# Multi Exit Configuration of Mesoscopic Pedestrian Simulation

Allan Lao
Lorma Colleges
San Fernando City
La Union
alao@lorma.edu

Kardi Teknomo
Ateneo de Manila University
Loyola Heights
Quezon City
teknomo@gmail.com

## ABSTRACT
A mesoscopic approach to modeling pedestrian simulation with multiple exits is proposed in this paper. A floor field based on Q-learning Algorithm is used. Attractiveness of exits to pedestrian typically is based on shortest path. However, several factors may influence pedestrian choice of exits. Scenarios with multiple exits are presented and effect of Q-learning rewards system on navigation is investigated.

## General Terms
Algorithms

## 1. INTRODUCTION
Pedestrian dynamics have been studied and presented in numerous papers. Models are mainly based on three methods namely, continuum model, social force model, and cellular automaton (CA) model.

In addition, most models are either macroscopic or microscopic. A macroscopic model minimizes variables and parameters and focuses mainly on the overall performance of pedestrian flow. In effect, macroscopic models reduce computational complexity. In contrast, microscopic models focus on the behavior of individuals. These models are able to reproduce individual properties such as acceleration, velocity, trajectory and interaction. However, there is a trade-off for performance since microscopic models are computationally expensive.

A hybrid of the two models is the mesoscopic model. Pedestrians are modeled as a group based on some commonalities such as same departure time or same route or simply walking in the same speed. Mesoscopic model improves performance without losing much pedestrian information. In this paper, we will use this approach.

Pedestrian simulations are widely used in areas such as evacuation dynamics, pedestrian flow performance and even store location evaluation in malls. Many of the environments used in these models deal with multiple exits and destinations. Hence, in this paper, we present a model to configure multiple sinks in a mesoscopic approach.

## 2. RELATED LITERATURE

### 2.1 Pedestrian Mesoscopic Models
There are only a few models available that aggregates pedestrian simulation into mesoscopic level. Many of these consider a group of pedestrians as a single entity thus losing much of the individual pedestrian information.

Florian et al (2001)[2] ,Hanisch et al. (2003)[3] and Tolujew and Alcalá (2004)[4], presented early papers that belongs to this group. Tolujew et. al(2004)[4] describe a framework for online control systems. The mesoscopic approach monitors group of pedestrians that show similar behavior because they have the same intention such as boarding a train.

Teknomo et al.(2008) [4] described mesoscopic model as a model where focus is not on single pedestrian interaction but on more aggregation of several pedestrians in a region. The pedestrian interaction between agents is presented as aggregation model of speed-density functions over space. To allow higher aggregation level, mesoscopic cells have a dimension of 1m x 1m and no upper bound but suggested a realistic limit of 3m x 3m. This is important not to blow up too much the size of walls, which becomes highly unrealistic. Several pedestrians may occupy the same cell at the same time with cell capacity having linear relation with the cell diameter. In contrast to the argument of [4] and [3], that mesoscopic level means group of pedestrians as an entity that move together, [6] modeled pedestrians as an individual agent. This allows measurement of flow performances derived from individual trajectory of agents.

### 2.2 Floor Field
Schadschneider (2001)[7] presented a concept of a floor field which acts as a substitute for pedestrian intelligence and leads to collective phenomena. This floor field makes it possible to translate spatial long-ranged interactions into non-local interactions in time. At the same time, pedestrians leave a trail similar to chemotaxis which affects the local interaction.

Floor fields have been presented in several forms. Varas et. al (2007)[12] describe floor field as a rectangular grid with each cell assigned a constant value representing distance to the exit. Lower value of cell directs the pedestrian to exits. High weights are assigned to walls to ensure that pedestrians do not move towards them while a value of 1 is assigned to exits.

The concept of Sink Propagation Value was presented by [6] which describes a set of monotonically increasing values that contains global information for navigation of pedestrian agents. Several methods were used like Smoothing Relaxation, Bellman Flooding, Distance Transform as well as Q-learning can be used to compute the SPV.

## 3. METHODOLOGY

### 3.1 Description of the model

The mesoscopic model is an extension of Teknomo e.t al. (2008)[6]. Is it based on the principle of Permission, Interaction and Navigation. Although the same framework was adopted, we implemented our own concepts in several aspects of the framework more specifically in navigation. There are 3 matrices used in the model each of which is discussed in separate sections. First is the layout matrix **L** which contains information on permission to move from cell to cell. Next is the density matrix **D** that represents number of agents currently in each cell. Last is the navigation matrix **N** which directs the agents towards one of the sinks.

*3.1.1 Pedestrian Movement*

For every time step, agents are evaluated of their next possible movement. If an agent can move, then one of the Moore neighborhood of 8 is considered as next the destination cell based on the computation of probability to enter the cell $Np$. Otherwise the agents stays on the same cell. Agent will move out of the cell only if

$$t \geq t_{in\ +s/u}$$

where $s$ is the speed and $u$ is the average speed of the current cell. This demonstrates the short travel time of the agent within the cell.

If an agent can move out of a cell, candidate cells considered for the next move is based on two functions, permission lookup and probability to enter cell. Since a cell can be bounded by walls, only neighbors with link are included in the list of choices. The probability to enter a cell is then computed as

$$Np = N_{Density}\ x\ N_{Navigation}$$

Since navigation matrix derived from Q-learning is not smooth, lateral movements are given priority in case of tie among top cell choices.

The next cell movement is defined as

$$Vt = argmax\ Np$$

*3.1.2 Layout Matrix*

The layout of the environment is modeled as a network graph formed as a regular lattice grid. Each vertex represents a discrete square cell. The edges indicate the presence or absence of wall that separates each cell. Unlike in previous studies, where walls are contained within a cell, our model makes use of edges to represent walls. Similarly, obstructions can be easily modeled as a cell surrounded by walls.

For every cell in the grid, a numerical equivalent is assigned to indicate position of walls within its four sides. A look up table is generated based on a 4-bit representation of a cell. The 4 bits represent four sides of the cell as shown in

| TOP | RIGHT | BOTTOM | LEFT |
|---|---|---|---|
| 1 | 1 | 1 | 1 |

**Figure 1. 4-bit representation of a cell.**

So for a cell with a value of 7 means that only the top side of the cell is open and the rest are closed.

Similarly, a movement guide indicates possible directions that an agent may take when navigating the environment. Each cell contains the predetermined list of possible movements. (see Figure 2)

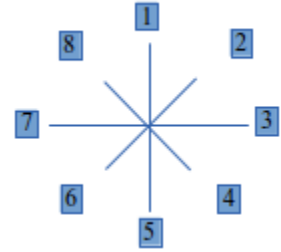

**Figure 2. Allowed pedestrian movement**

*3.1.3 Pedestrian Interaction*

The interaction among pedestrians is represented by a function of cell density. Two functions of density are imposed on each cell. First is the speed density relationship and second is the probability to enter the next cell. For simplicity, a look up table is used to determine the speed-density relationship. A detailed explanation of speed-density can be found on [6][8]. Table 1. shows the relationship between cell density, speed and probability to enter next cell.

**Table 1. Speed-Density relationship and probability to enter next cell based on a 1x1 meter cell**

| Cell Density | Speed in m/s | Probability to enter cell |
|---|---|---|
| 0 | 1.44 | 1.0 |
| 1 | 1.12 | 0.8 |
| 2 | 0.84 | 0.6 |
| 3 | 0.56 | 0.4 |
| 4 | 0.28 | 0.2 |
| 5 | 0.00 | 0.0 |

Each cell has a uniform diameter computed as the average between the diameter of the inner circle and the outer circle. Depending on the speed and density, there is a short travel time within the cell. Pedestrian movement is assumed to be uniform inside the cell. Speed of the pedestrian is dependent on the density of the cell as shown in Table 1. Each pedestrian will only have a single origin cell and a single destination cell. Obviously, the origin cell must not be the same cell as the destination cell. Speed and density is computed for every time step.

### 3.1.4 Pedestrian Navigation

The third component of the PIN model is the navigation matrix which gives direction to pedestrian agents. The principle behind the navigation matrix is simple; a series of monotonically increasing numbers is embedded in each cell such that destination sinks have high values and is propagated across the environment. It is similar to the concept of a signal from a transmitter such that when the agent moves it follows the location which emits a stronger signal and is eventually directed towards the origin of the signal.

The monotonically increasing number is known as Sink Propagation Value. [6][9] describes SPV as a function assigning a value to each vertex that is implementing a general notion of distance from the sink. It is a value on vertex that monotonically increasing (or decreasing) from sink vertex according to the minimum distance function from that vertex to the sink. Therefore, SPV allows global information to be stored locally on the vertices. [6][9] describes the properties of SPV in detail.

The quality of SPV generated is crucial to the behavior of pedestrian agents. The agent may eventually find its destination but have traversed unusual routes like areas close to walls. Also, agents may exude a military like behavior moving mostly in lateral directions instead of a smooth curve towards destination. Hence, the design of the SPV must be carefully examined such that SPV results form a radial like wave from sink nodes outwards.

There are several ways to compute for SPV such as Smoothing Relaxation, Bellman-Ford Algorithm, A*, Distance Transform and Q-learning Algortihm. In this paper we used a dual of Q-learning[9].

## 3.2 Configuring Exits

There are several points of interest in a navigation field which we call here as basins. Teknmo(2008)[9] categorized basins as 1) source basin, 2) sink basin 3) source and sink basin 4) saddle. Source basin is where agents originate. Sink basin like exit door is the final destination and agents will eventually disappear. Saddle basin, is a temporary destination like a cashier or a shop where a service time is consumed before an agent moves to other destination. In this paper, focus is towards assigning weights on basins.

Exits have varying degree of attractiveness based on individual preferences and goals of pedestrians. A female comfort room is of course attractive to female pedestrians and a ramp is certainly better for persons with disabilities. Although we have individual preferences, many of these can be generalized like an escalator is preferred more than a stair for most pedestrians. Similarly main exit is preferred more than the emergency exit.

Most floor fields or navigation fields are based on distance metrics[7][10][12]. It implies that the closer the exit to the source the higher the probability of choosing that exit route. Figure 3 shows a floor field for a symmetrical layout with multiple exits and with same attraction value for all exits.

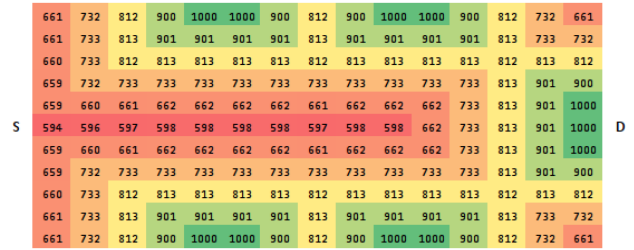

**Figure 3. Five exits with equal strengths**

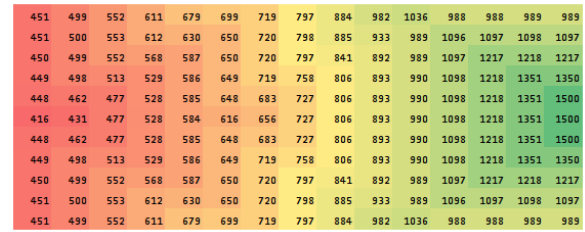

**Figure 4. Five exits with main exit strength increased 3x**

Clearly, exits closer to the source have the higher probability of being chosen as shown in Figure 3. However, if we adjust weights such that exits are unequal, the other exit influences the other. An example is shown in Figure 4 with the weight of the main exit multiplied 3 times. This resulted in overpowering the rest of the exits ahead of the main exit.

### 3.2.1 Q-learning

Configuration of several exits in a layout depends primarily in the reward system of the Q-learning algorithm. Exits closer to the source are by default more attractive than exits at a distance. Increasing the pulling power of exits allow pedestrians to navigate towards the exit's direction. Similarly, decreasing the exit value would shift navigation direction towards exit with higher value.

A subset of reinforcement learning algorithm, Q Learning is a general computational approach to learning from rewards and punishment [13]. The algorithm works by maintaining a list of possible actions from a given state and assign Q-values to them. For every step in the algorithm, it selects one possible action and observe the rewards it collects from the environment. Q-values are updated based on simple rules. Thus, after a finite number of steps, the algorithm learns the value cost from all actions [14]. From there, the algorithm can select the optimal path from source to destination.

Although Q-learning algorithm is certainly not the fastest in the family of searching and path finding algorithms, its reward and punishment mechanism is a powerful tool to rationalize the environment. In addition, the rate at which Q-learning algorithm learns the environment is not critical since the process is performed only once during the design of the model and not at run time.

The process of generating a navigation matrix is fairly simple. Given a layout matrix **L,** rewards matrix **R** is derived. The **R** matrix is an adjacency matrix of size $\langle m \: x \: n \rangle \: x \: \langle m \: x \: n \rangle$ where $\langle m \: x \: n \rangle$ is the size of the layout matrix. Rewards matrix contains link information between cells and corresponding reward for visiting the cell. The physical layout can be represented as an undirected network graph as shown n Figures 5 and 6.

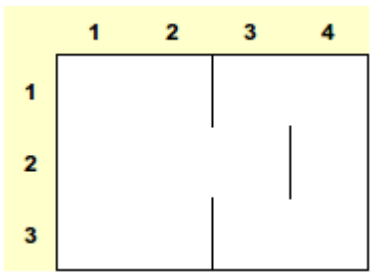

**Figure 5. Physical Layout**

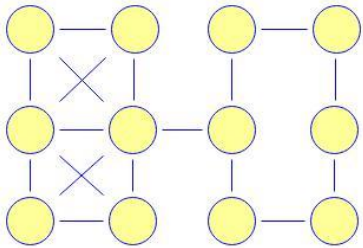

**Figure 6. Network Graph**

By following the Q-learning Algorithm from the tutorial [8] we get the Q matrix and subsequently, by getting the diagonal values we get the navigation matrix **N**.

## 4. SIMULATION RESULTS

### 4.1 Scenario 1 : Symmetrical 3 exit layout.

In this scenario we simulate a 20m x 30m layout with a main exit and 2 side exits. Size of mesoscopic cell is 1m x 1m. The layout is inspired by a mall in San Fernando City where 3 cinemas are located on the left side of the layout and the main exit opposite the cinemas. The two scenarios 1 and 2 are identical except that in scenario 2, the sink value of the main exit is reduced by 50%. Movie house doors have widths of 4 meters. Side exit doors both have 2 meters width while the main exit is 4 meters wide. The main exit is given a high sink value to pull agents from the movie house towards the main door (as seen in Figure 5).

After last full show, the mall is almost empty except for the moviegoers rushing to go out of the mall. Both scenarios were simulated with 60 pedestrians at peak and time is measured in seconds.

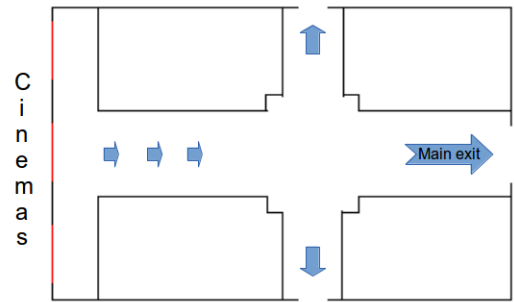

**Figure 7. Scenario 1**

| Time | Scenario 1 (a) | Scenario 1(b) |
|---|---|---|
| 1 | | |
| 5 | | |
| 10 | | |
| 15 | | |
| 20 | | |
| 25 | | |

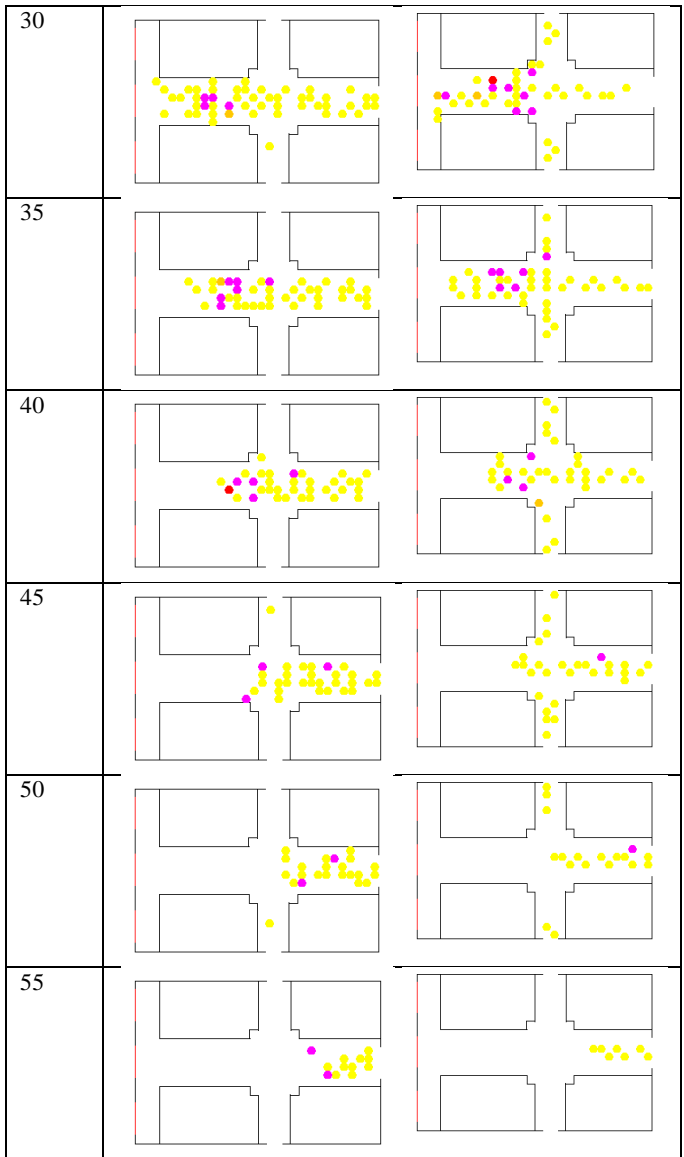

**Figure 8. Movement pattern of 3 exit layout.**

The first scenario where a high sink value was assigned for the main exit, pedestrians tend to move within the middle of the main route. The movement pattern shows that most of the pedestrians walked towards the main exit while only a handful were attracted to the nearer side exits. Those that went to the side exits were actually walking at the outskirt of the crowd or were either pushed or left out during pedestrian interaction.

The second scenario in which the sink value of the main exit is reduced by 50%, it shows that the pedestrians were actually divided between the three exits with still the main exit getting the most. The reduction caused the sink to lose pulling power. Both scenarios exhibited phenomenon such as crowding, wall avoiding and route choice.

## 4.2 Scenario 2 : Symmetrical 3 exit layout with elevator and stair.

We extended the first scenario to include an elevator alongside a stair close to the main exit. The purpose is to demonstrate the capability of the model to designate a main path other than the wider and with equal distance to the main exit. Figure 9 show the floor layout.

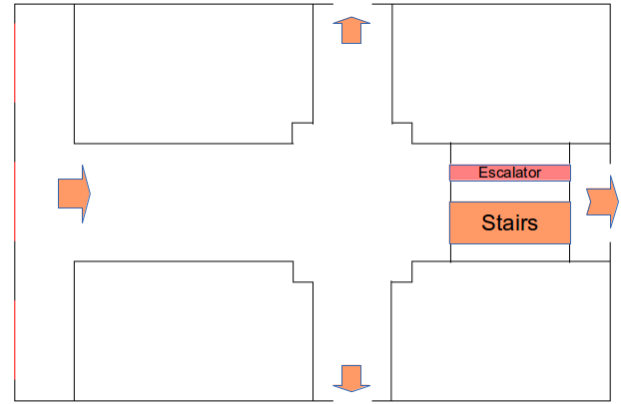

**Figure 9. Layout with escalator and stair**

In this layout, we designed the escalator to be the preferred choice of pedestrians. Figure 9. shows a portion of the floor field for the said scenario. Noticeably, the escalator has higher values within its approach whereas the stair is affected by the attractiveness of the main exit.

| 18 | 2500.7 | 2777.6 | 3084.1 | 3084 | 18 | 19 | 19 | 19 |
|---|---|---|---|---|---|---|---|---|
| .1 | 2502.4 | 2778.4 | 3085 | 3425.6 | 3425.5 | 18 | 18 | 18 |
| .7 | 2503.1 | 2779.1 | 3085.8 | 3426.5 | 3805.2 | 4224 | 4225.8 | 18 |
| .2 | 2503.7 | 2779.8 | 3086.5 | 3427.3 | 3806.1 | 4226.8 | 4694.4 | 5214.9 |
| .7 | 2503.2 | 2779.2 | 3085.9 | 3426.6 | 3805.2 | 4225.9 | 4225.8 | 18 |
| .2 | 2502.6 | 2778.5 | 3085.1 | 3425.8 | 3804.3 | 3885.5 | 4315.1 | 4791.4 |
| .6 | 2501.9 | 2777.8 | 3084.3 | 3424.9 | 3498.8 | 3885.5 | 4315.1 | 4791.4 |
| 53 | 2501.2 | 2777 | 3083.4 | 3150 | 3497.9 | 3882.7 | 3884.5 | 18 |
| .2 | 2500.4 | 2776.1 | 2836.1 | 3149.1 | 3149 | 18 | 18 | 18 |
| 18 | 2497.6 | 2553.6 | 2835.2 | 2835.1 | 18 | 19 | 19 | 19 |

**Figure 10. Portion of floor field leading to the escalator**

Even if the width of the stair is twice that of the escalator and distance to the main exit is the same, we were able to assign higher weight to the escalator. Note that in this paper, velocity was not considered for both escalator and stair since the goal of the paper is to provide the right floor field.

It is interesting to note that pedestrians tend to form a queue within the approach of the escalator. Some pedestrians choose to walk down the stairs to avoid the queue. The rest of the pedestrians close to the wall of the walkway have higher probability of exiting towards the side. Figure 11 shows a snapshot of time step 45 of the simulation. Yellow dots represent a single pedestrian in a cell while dots with darker colors means more than 1 pedestrian is currently occupying the cell.

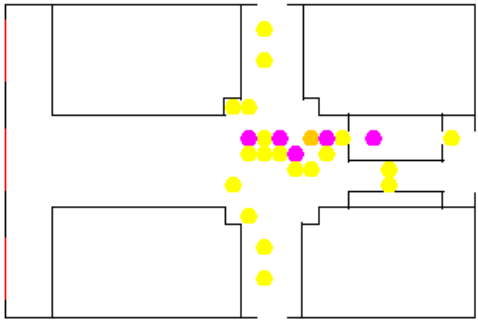

**Figure 11. Pedestrians at time step 45**

## 4.3 Comparison of mesoscopic and microscopic model.

A simple comparison on the performance of mesoscopic model against a microscopic model is presented here. Note however that it is not the objective of this paper to investigate and appraise performance results between the two models. For purposes of presentation, a simple scenario of 10m x 15m layout is modeled here. An equivalent 20x30 grid layout represents the microscopic model. The former has a cell dimension of 1m x 1m and the latter with a 0.5m x 0.5m cell dimension, hence the grid size of microscopic is twice the dimension of mesoscopic model.

In addition, the mesoscopic model can have as many as four pedestrians occupying a single cell at one point in time. Figure 12 show 30 pedestrians at simulation time step 17. The delta time used is 0.50 seconds per time step.

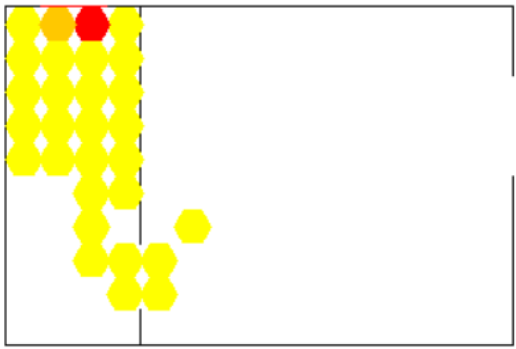

**Figure 12. Mesoscopic model at time step 17**

Microscopic models on the hand, allows only a single pedestrian per cell, where the dimension of a cell closely resembles the body size of an average pedestrian which in this case is 0.5m x 0.5m.

Figure 13 shows 30 pedestrians at time step 35 with delta time of 0.5 seconds per simulation step.

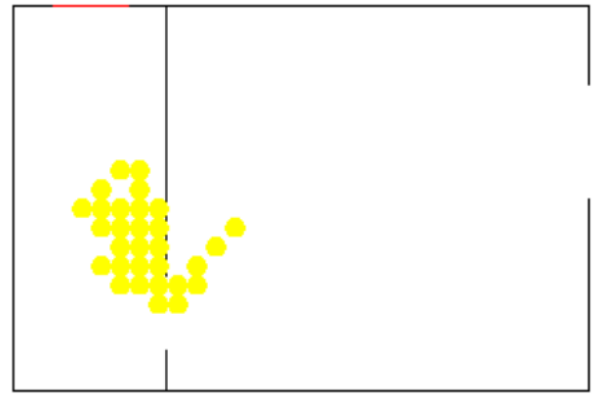

**Figure 13. Microscopic model at time step 35.**

A total of 50 simulation runs covering 1 to 50 agents is done on both models. Two basic performances were measured that is, average travel time and average distance travelled. Figure 14 and 15 shows performances of the two models.

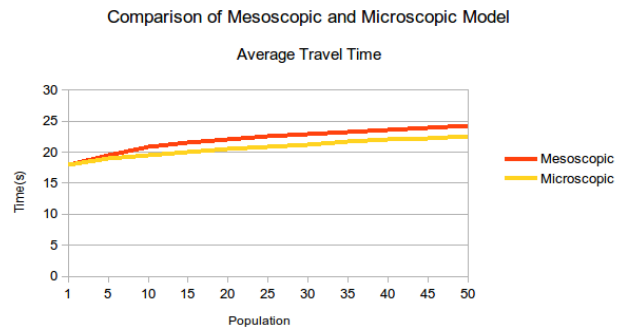

**Figure 14. Average travel time**

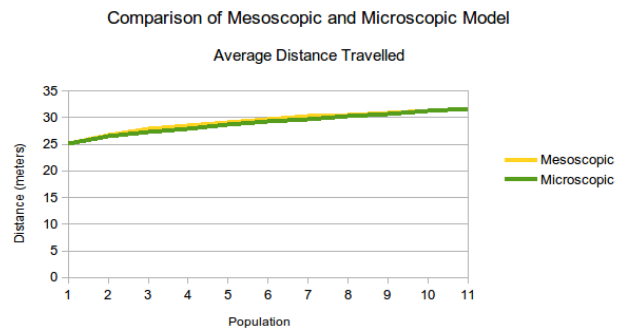

**Figure 15. Average distance travelled**

The average travel time shows that both models are almost identical. The slight difference in the average travel time can be attributed to the speed-density look up table where speed is rounded off to two decimal places. It is also interesting to note that the performances of a single pedestrian in both models are the same.

The average distance travelled showed even a closer figure with both models showing increasing distance as population increases. This can also be attributed to interaction among pedestrians.

## 5. CONCLUSIONS

In this study, a mesoscopic approach to modeling pedestrian simulation with multiple exits is presented. We combine Q-learning Algorithm and simple rules of Cellular Automata to design floor fields and pedestrian interaction. The model presented can be a valuable tool in simulating evacuation scenarios with multiple exits as well as pedestrian flow analysis for different routes.

## 6. RECOMMENDATIONS

An analysis of the relationship of distance and the weight assigned to destination sinks should be investigated to determine the optimal weight value to be assigned per exit area. This is necessary to avoid trial and error in the determination of weight values.

## 7. ACKNOWLEDGMENTS

This study is supported in part by the Commission on Higher Education and Lorma Colleges.